\documentclass[onecolumn]{elsart}
\usepackage{exscale}
\usepackage{amssymb,amsbsy}
\usepackage[dvips]{graphicx}

\begin{document}
\begin{frontmatter}
\title{$\boldsymbol\omega$ attenuation in nuclei}
\thanks{We gratefully acknowledge support by the Frankfurt Center for Scientific Computing
and the DFG within the scope of the SFB/TR16.}
\author{P.~M\"uhlich} and
\author{U.~Mosel}
\address{Institut f\"ur Theoretische Physik, Universit\"at
Giessen, D--35392 Giessen, Germany}
\begin{abstract}
We study inclusive $\omega$ photoproduction in nuclei and
propose a measurement of the nuclear transparency ratio as a means
to learn about the in-medium properties of the $\omega$ meson. To
this end we are using the semi-classical coupled-channel BUU
transport approach that allows for a consistent treatment of all
nuclear effects. The conditions of our calculations are chosen such
as to match the setup of existing experimental facilities at ELSA or
MAMI C. We show that the observables (total production cross section
and transparency) indeed are sensitive to the $\omega N$ interaction
in the nuclear medium.
\end{abstract}
\end{frontmatter}

\section{Introduction}

One intriguing theme of current nuclear physics research is the
change of the vector meson properties once they are embedded in a
strongly interacting environment. Experimental hints towards such
changes -- albeit partly under much debate -- have been obtained
from heavy ion collisions
\cite{Agakichiev:2005ai,Usai:2005zh,Adams:2003cc} as well as from
photon- \cite{Trnka:2005ey} and proton-induced \cite{Ozawa:2000iw}
reactions on finite nuclear targets. All information about the
intrinsic properties of a vector meson are encoded in its spectral
function. The in-medium spectral densities of the vector mesons are
connected to the vector meson-nucleon and -- in the case of finite
temperatures -- to the vector meson-pion interaction at finite
$\rho$ and/or $T$. Thus, they depend on information that cannot be
obtained from elementary scattering processes. Whether vector mesons are
-- in addition -- also affected by QCD condensates and their
in-medium changes \cite{Brown:1991kk,Hatsuda:1991ez,Leupold} is still a
matter of debate.

In the present paper we concentrate on the spectral density in the
vector-isoscalar channel. Theoretical results for the $\omega$
in-medium spectral function have been obtained on the basis of
rather different methods
\cite{Bernard:1988db,Klingl:1997kf,Klingl:1998zj,Post:2000rf,Lutz:2001mi}.
The outcome of these models covers a large area in the mass/width
plane, ranging for the mass from the free $\omega$ pole at $m_0=782$
MeV down to roughly 640 MeV for an $\omega$ meson at rest in nuclear
matter at saturation density $\rho_0$. For the width, that amounts
to $\Gamma_0=8.4$ MeV in vacuum, in-medium values of up to about 70
MeV at $\rho_0$ have been obtained. The origin of the $\omega$ medium
modifications at finite density and zero temperature have for
instance been attributed to the collective excitation of
resonance-hole loops \cite{Post:2000rf,Lutz:2001mi} that lead to
additional structures in the spectral function. Another source of
medium modifications can be the renormalization of the pion cloud,
considered in \cite{Klingl:1997kf,Klingl:1998zj}, that results in a
rather drastic shift of spectral strength to the low mass region.

One photoproduction experiment with the aim to learn about the
$\omega$ in-medium spectrum has been done only recently at ELSA
\cite{Trnka:2005ey}. $\pi^0\gamma$ pairs from $\omega$ mesons
photoproduced in finite nuclear systems have been detected in order
to reconstruct the invariant mass spectrum of the decayed $\omega$
mesons. The $\pi^0\gamma$ decay mode is particularly sensitive to
the vector-isoscalar channel as the anomalous coupling
$\omega-\pi^0\gamma$ is rather large \cite{Klingl:1996by}. In a
preceding paper \cite{Muhlich:2003tj} we have shown that despite the
strong final state interactions of the outgoing pion it is possible
to obtain information on the $\omega$ in-medium spectrum from such a
measurement. Whereas the authors of \cite{Trnka:2005ey} report a
downward shift of the $\omega$ in-medium peak, the assignment of an
in-medium width was not possible due to the only moderate detector
resolution. Even cleaner signals for the in-medium change of the
vector meson properties are expected from an observation of the
dilepton yield as this final state is essentially free of any final
state interactions. The analysis of such an experiment done at JLAB
is presently under way \cite{Djalali:2004}.

An alternative method to study the $\omega$ self energy in nuclei is
an attenuation measurement of the $\omega$ flux in $\omega$
photoproduction off nuclei. Already in early experiments on the
$\rho$ meson properties in nuclei \cite{Alvensleben:1970uw} this
method has been used to yield the first extraction of the $\rho N$
cross section from photoproduction experiments on nuclei. More
recently, the authors of \cite{Cabrera:2003wb} haven taken this up
for the case of $\phi$ photoproduction. In \cite{Muehlich:2005kf} we
have shown that the $A$-dependence of the total $\phi$ meson yield
is indeed sensitive to the $\phi$ width in the nuclear medium. Data
taken at SPring8/Osaka \cite{Ishikawa:2005aw} show the need for an
explicitly higher $\phi$ absorption cross section as compared to
usual quark model estimates in order to obtain agreement between
theory and experiment. In the present paper we take over the same
idea in order to show the possibility to obtain information on the
$\omega N$ interaction in nuclei via a measurement of the
$\pi^0\gamma$ yield from photon-nucleus reactions. Such experiments
are presently being evaluated.

This article is structured in the following way:
In the following section we introduce the semi-classical framework 
that we use to model inclusive particle production in photon nucleus reactions.
In section \ref{Results} we discuss our results that are
structured according to different scenarios of medium modifications.
Finally we turn to the discussion of $\omega$ photoproduction in nuclei
within the semi-analytic Glauber model
and show its power in analyzing our previously obtained transport results.
We close with a summary in section \ref{Summary}.


\section{Model}\label{Model}
\subsection{Transport Model}

In order to simulate the
production and propagation of $\omega$ mesons in finite nuclear
targets in an as realistic way as possible we employ the
semi-classical Giessen BUU transport approach. This approach
describes incoherent photon-nucleus reactions and has successfully
been applied previously to the study of medium modifications of
the $\rho$, $\omega$ and $\phi$ mesons by means of $e^+e^-$
\cite{Effenberger:1999ay}, $\pi^0\gamma$ \cite{Muhlich:2003tj} and
$K^+K^-$ \cite{Muhlich:2002tu,Muehlich:2005kf} photoproduction in
nuclei. The model aims at a complete description of all nuclear
effects, i.~e. Fermi motion, Pauli blocking, nuclear shadowing,
strong and electromagnetic potentials and elastic and inelastic
scattering processes including sidefeeding and regeneration of all
propagated particles. Moreover, not only the elastic, but also
inelastic production channels, i.~e. $\gamma N\rightarrow
VX~(X=\pi N,\pi\Delta,...)$, are included by means of nucleon
resonance decay channels or the LUND
model FRITIOF \cite{Andersson:1992iq}. A description of the
Giessen BUU model can be found in
Refs.~\cite{Effenberger:1999ay,Muhlich:2002tu,Teis:1996kx,Lehr:1999zr,Falter:2002jc}
and references therein. In \cite{Muehlich:2005kf} we have
demonstrated the necessity to carefully account for rescattering
processes and hadronic potentials in order to reliably extract the
$\phi N$ interaction strength from $\phi$ photoproduction experiments.

\begin{figure}
\begin{center}
\includegraphics[scale=1.]{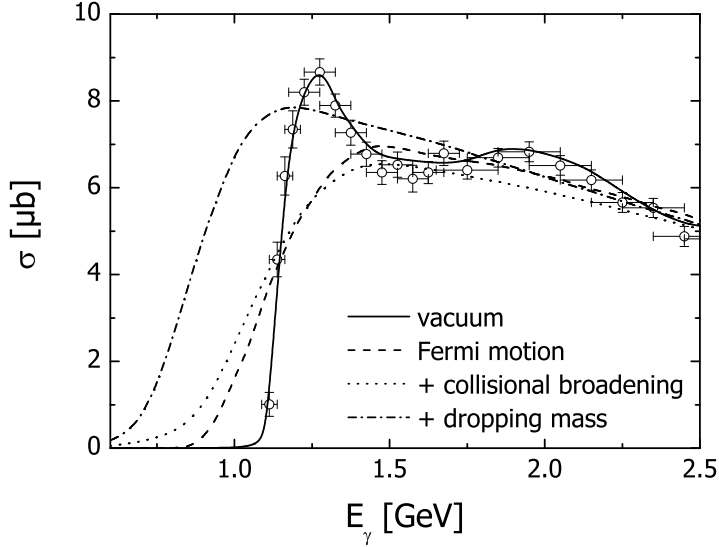}
\caption{Total exclusive $\omega$ photoproduction cross section on the nucleon.
For the dashed curve the effect of Fermi motion at normal nuclear matter density 
$\rho_0$ has been taken into account. The dotted and dash-dotted curves in 
addition include medium modifications according to Eqs.~(\ref{gcoll}),~(\ref{mass}).
The data points are taken from \cite{Barth:2003}.}
\label{Fig_proton}
\end{center}
\end{figure}

The $\omega$ photoproduction cross section at finite density we
calculate by allowing the $\omega$ meson to take arbitrary masses
and convoluting the obtained expression with the $\omega$ spectral
function at given density. The matrix element we calculate within a
tree-level model, including the contribution from the $t$-channel
$\pi$-exchange as well as the $s$- and $u$-channel contributions
including the nucleon and the $P_{11}(1710)$ as intermediate states
following the findings of \cite{Penner:2002md}. In vacuum, this
model describes the experimentally determined angular distributions
very well. The more complicated structures in the total cross
section \cite{Barth:2003} are, however, hard to obtain within such a
simple approach. For the total production rate we therefore use the
measured cross section as obtained by the SAPHIR collaboration
\cite{Barth:2003}. The implementation of this experimental
information into our transport simulation is explained in
\cite{Muhlich:2003tj}. There also the inclusion of inclusive
photoproduction channels is discussed. Due to the lack of
experimental information we assume the cross section from neutrons
to be the same as the cross sections from protons. In
Fig.~\ref{Fig_proton} the exclusive cross sections from the proton
target in vacuum as well as at non-zero density, assuming different
scenarios of medium modifications of the $\omega$ that will be
subject to the following discussions, are shown.

\subsection{The $\omega N$ interaction}

In the very same spirit as in a preceding paper for the case of
$\phi$ photoproduction \cite{Muehlich:2005kf} we use
parametrization for the $\omega N$ total cross sections. For
$\omega N$  elastic scattering we have
\begin{eqnarray}\label{sel}
\sigma_{\mathrm{el}}=\left[5.4+10 \exp\left(-0.6~|{\bf q}|\right)\right]~\mathrm{mb}
\end{eqnarray}
where ${\bf q}$ is the laboratory momentum of the $\omega$ meson
in GeV. This expression has been obtained in \cite{Lykasov:1998ma}
by an interpolation of the low energy cross section, calculated
from a microscopic model, and the high energy limit obtained
within an additive quark model. Also for the inelastic cross
section we use the parametrization from \cite{Lykasov:1998ma}
\begin{eqnarray}   \label{sinel}
\sigma_{\mathrm{in}}=\left[20+\frac{4}{|{\bf q}|}\right]~\mathrm{mb},
\end{eqnarray}
that again interpolates the low energy cross section, given by the
sum of the individual contributions $\omega N\rightarrow \pi N,2\pi
N,\sigma N,\rho N$ and $\rho\pi N$, and the high energy limit,
estimated within the strict vector meson dominance model (SVMD). For
high energies ($\sqrt{s}>2.2$ GeV) inelastic scattering events are
simulated within the FRITIOF model \cite{Andersson:1992iq}. Both the
cross sections (\ref{sel}) and (\ref{sinel}) are obviously only
estimates. However, more recent coupled-channel analysis of pion-
and photon-induced $\omega$ production cross sections on the nucleon
\cite{Shklyar:2004ba} yield results that are comparable in magnitude
with those of \cite{Lykasov:1998ma} although they fall of with
momentum more quickly than (\ref{sinel}). Ultimately, attenuation
experiments can help to determine at least the inelastic cross
section.

The collisional width of the $\omega$ we obtain via the low density
theorem within a local density approximation
\begin{eqnarray}\label{gcoll}
\Gamma_{\rm coll}(q_0=\sqrt{m_V^2+{\bf q}^2},{\bf
q};\rho(\vec{r})) = \frac{4}{m_V} \int\frac{d^3p}{(2\pi)^3}
\Theta(|{\bf p}|-p_F(\vec{r}))\nonumber\\
\times\frac{k\sqrt{s}}{E_N({\bf p})}
\sigma_{VN}(s)\,
\end{eqnarray}
with $k$ being the center-of-mass momentum of nucleon and
$\omega$, $E_N$ the nucleon on-shell energy in the laboratory
frame and $p_F(\vec{r})$ the local Fermi momentum. The cross
section $\sigma_{VN}$ is the total $\omega N$ cross
section containing all quasi elastic and absorption channels. For
$\omega$ mesons at rest we find a collisional width of 37 MeV at
normal nuclear matter density, a value that lies within the range
of most of the more elaborate approaches
\cite{Klingl:1998zj,Post:2000rf,Lutz:2001mi}. We note,
however, that we assume in our calculations a Breit-Wigner shape
for the spectral function of the $\omega$ and thus do not allow
for a multi-humped structure as obtained in
\cite{Post:2000rf,Lutz:2001mi}.


\begin{figure}
\begin{center}
\includegraphics[scale=1.]{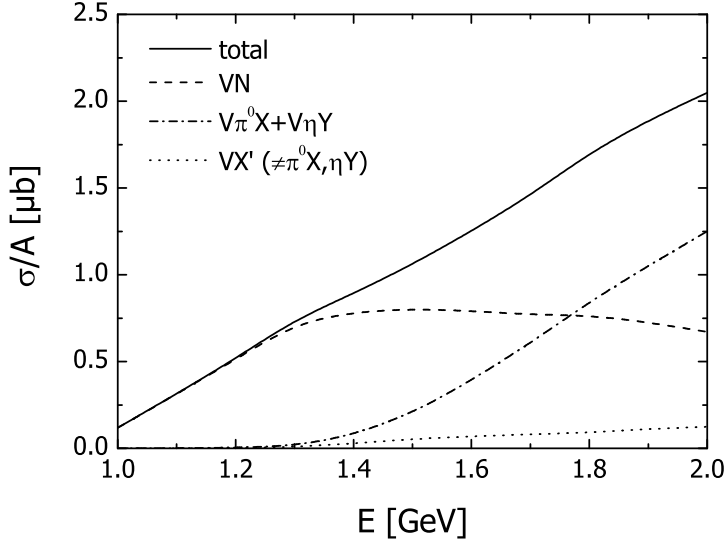}
\caption{Total inclusive $\omega\rightarrow\pi^0\gamma$ photoproduction cross section
from \nuc{40}{Ca} \emph{without} final state interactions and
Pauli blocking. The dash-dotted line gives the cross section for
the production of inclusive final states including a $\pi^0$ or an
$\eta$ meson. The dashed line, labeled $VN$, gives the 
$\gamma A\rightarrow\omega A\rightarrow\pi^0\gamma A$ cross section.}\label{Fig_sigvx}
\end{center}
\end{figure}

\section{Results}\label{Results}
\subsection{Observables}

As a measure for the $\omega$ width in nuclei we use the so-called
nuclear transparency ratio:
\begin{eqnarray}\label{ta}
T_A=\frac{\sigma_{\gamma A\rightarrow VX}}{A\sigma_{\gamma
N\rightarrow V X}}~,
\end{eqnarray}
i.~e. the ratio of the inclusive nuclear $\omega$ photoproduction cross
section divided by $A$ times the same quantity on a free nucleon.
It can be interpreted as the momentum- and position-space averaged
probability of an $\omega$ meson to get out of the nucleus. The loss of
flux is obviously related to the absorptive part of the $\omega$ nucleus
potential and thus to the $\omega$ width in the nuclear medium.

Moreover, the total nuclear production cross section and the 
transparency ratio can also be sensitive to
the real part of the $\omega$ nucleus potential. An attractive mass
shift, e.~g. of the Brown-Rho-type \cite{Brown:1991kk}
\begin{eqnarray}\label{mass}
m_V^*\equiv m_V^*({\bf r})=m_V^o\left[1-\alpha\frac{\rho({\bf
r})}{\rho_0}\right]~,
\end{eqnarray}
where $m_V^o=782$ MeV is the physical $\omega$ mass in vacuum,
causes an in-medium lowering of the $\omega N$ threshold due to the
smaller $\omega$ mass at finite density. This results in a divergent transparency
ratio at the vacuum threshold and an explicit enhancement in the
near threshold region. We note, however, that such a threshold
enhancement is not a unique signal for a lowering of the in-medium
$\omega$ mass but can also -- albeit weaker -- be generated by Fermi motion and coherent
nuclear $\omega$ production. We will discuss this issue in more detail
among the results of our calculations.


\subsection{Imaginary part of the $\omega$ nucleus potential}

In order to first explore the imaginary part of the $\omega$ nucleus
potential, we perform calculations at the fixed photon beam energy
of $E_{\gamma}=1.5$ GeV. On one hand, this energy is well above
the free production threshold ($\approx 1.1$ GeV) so that here one
is essentially free of any threshold effects, i.~e. medium
modifications of the elementary cross section for instance due to
the in-medium broadening of excited nucleon resonances. 
Also the sensitivity to a density-dependent shift of the $\omega$ 
pole mass becomes small
for this beam energy as we will show in the following section.
On the other hand, the chosen energy is low
enough so that any inclusive production channels $\gamma
N\rightarrow VX$ with $X\ne N$ (dotted line in Fig.~\ref{Fig_sigvx})
are of minor importance.
Moreover, events from the dominant inclusive channel $\gamma N\rightarrow V\pi^0 X$
can easily be suppressed experimentally.
Thus, ambiguities due to the elementary production process are minimized
for the chosen beam energy of 1.5 GeV.

\begin{figure}
\begin{center}
\includegraphics[scale=1.]{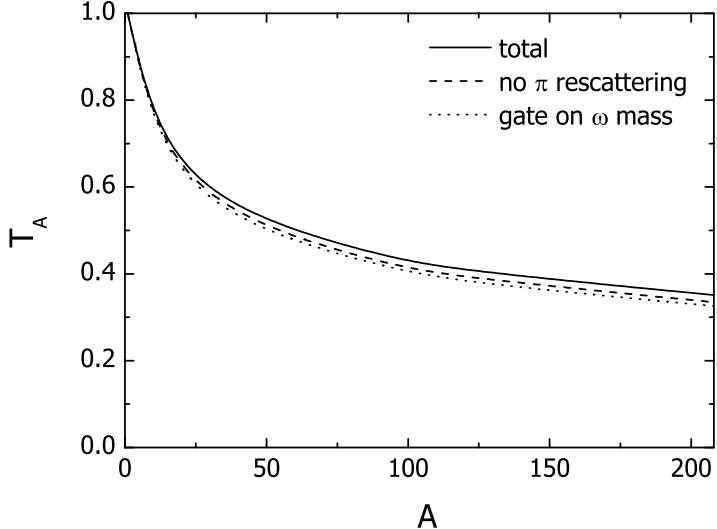}
\caption{Nuclear transparency ratio from BUU calculations
as function of the target mass number. The photon energy is
$E_{\gamma}=1.5$ GeV. The solid curve is calculated without any restrictions
on energy and momentum of the $\pi^0\gamma$ pair. For the dashed line
pions that interacted via a quasi elastic collision are removed from the
flux. For the dotted line in addition the $\pi^0\gamma$ invariant
mass has been restricted to $0.75~\mathrm{GeV}\le M\le0.81~\mathrm{GeV}$.}
\label{Fig_ratio1}
\end{center}
\end{figure}

The BUU calculations have been performed for the targets \nuc{12}{C}, \nuc{40}{Ca},
\nuc{93}{Nb}, \nuc{120}{Sn} and \nuc{208}{Pb} with an eye on the TAPS experiment
where the $\omega$ is detected via the $\pi^0\gamma$ decay channel. In Fig.~\ref{Fig_ratio1} we show our results
obtained within the standard scenario, i.~e. using the cross sections as given
by Eqs.~(\ref{sel}), (\ref{sinel}) and including collisional broadening
of the $\omega$ according to Eq.~(\ref{gcoll}) as a medium modification only.
In the experimental
analysis one tries to get rid of pions that rescattered in the
medium since these $\pi^0\gamma$ pairs essentially lose all information
about their source. This can be done easily as these $\pi^0\gamma$ pairs
appear at much lower values of the invariant mass and also the $\pi$ kinetic energy
is smaller as compared to pions from the $\omega$ decay that leave the
target nucleus untouched \cite{Messchendorp:2001pa,Muhlich:2003tj}.
Removing the pions that interacted via quasi
elastic collisions from the total flux, we obtain the dashed line
in Fig.~\ref{Fig_ratio1}, i.~e. only a small reduction of the transparency ratio
is observed. This becomes immediately clear if one realizes that most of the
$\omega$ mesons decay outside the nucleus and therefore the probability for
the pion to scatter from the target nucleons is small. If, in
addition, restrictions on the $\pi^0\gamma$ invariant mass are imposed in order
to gate on the $\omega$ decay component in the $\pi^0\gamma$ spectrum, only a slight
reduction can be observed. This is shown by the dashed curve in
Fig.~\ref{Fig_ratio1}, where the condition $0.75~\mathrm{GeV}\le M\le0.81~\mathrm{GeV}$
has been applied. Again the reason for this marginal effect is the
only tiny contribution of $\omega$ decays in the medium where the $\omega$ spectral
distribution becomes broad.

\subsection{Real part of the $\omega$ nucleus potential}

\begin{figure}
\begin{center}
\includegraphics[scale=1.2]{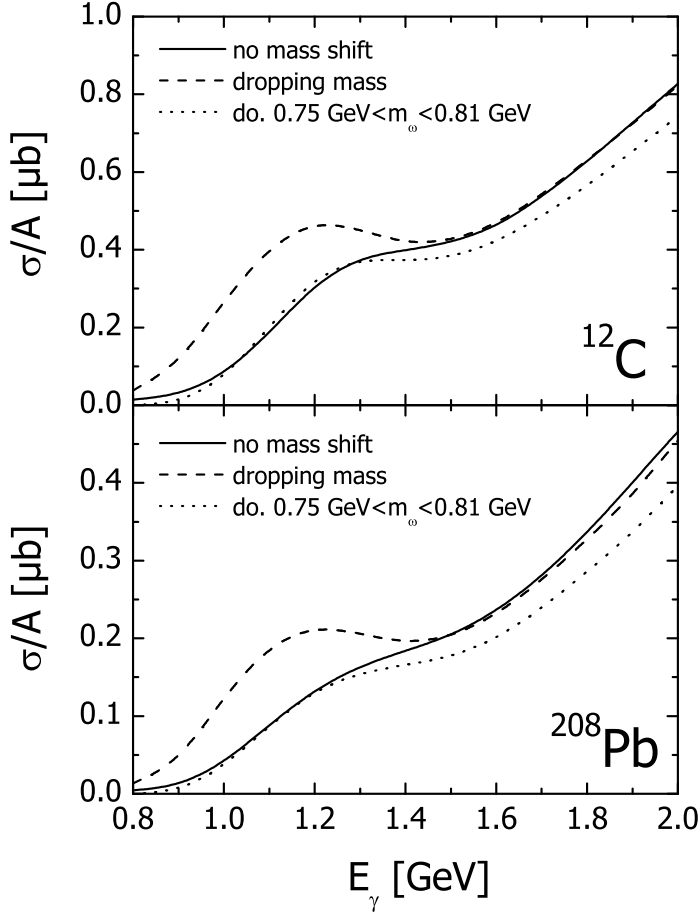}
\caption{Total $\omega$ photoproduction cross section from
\nuc{12}{C} and \nuc{208}{Pb} as function of the photon beam
energy. Results with and without a density dependent shift of
the in-medium $\omega$ are shown. The dotted curves are obtained
by imposing the condition that $0.75~\mathrm{GeV}\le M \le 0.81~\mathrm{GeV}$.}
\label{Fig_tsigma}
\end{center}
\end{figure}

\begin{figure}
\begin{center}
\includegraphics[scale=1.2]{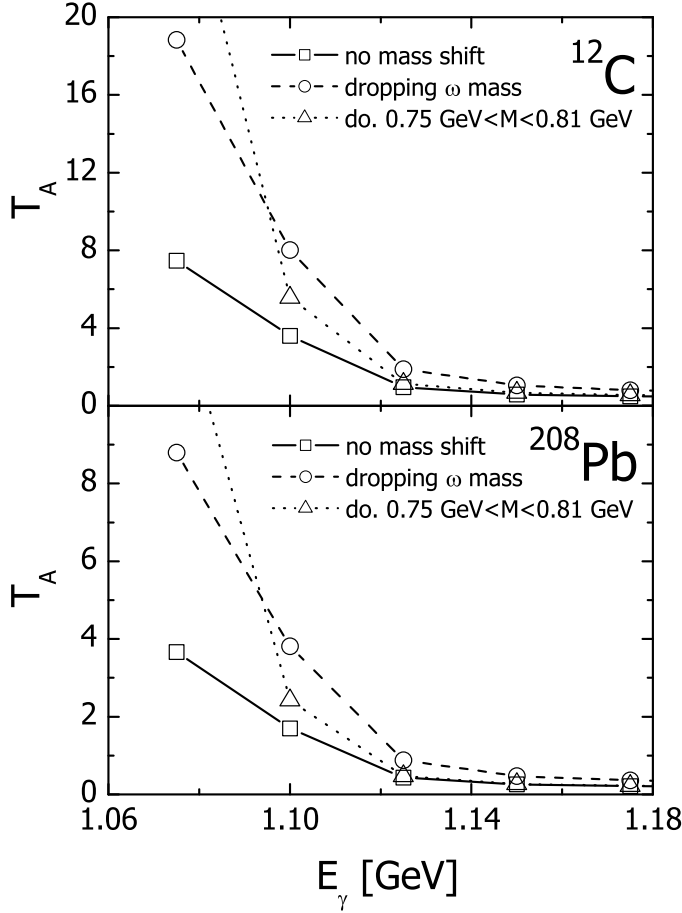}
\caption{Nuclear transparency ratio as function of the photon
energy in the threshold region. Results with and without a
density-dependent shift of the in-medium $\omega$ mass are shown.
The dotted curves are obtained by imposing the condition that
$0.75~\mathrm{GeV}\le M \le 0.81~\mathrm{GeV}$.}
\label{Fig_thres}
\end{center}
\end{figure}

In order to obtain also information on the real part of the $\omega$
nucleus potential, one has to examine the energy dependence of the
total cross section in the threshold region. This is shown in
Figs.~\ref{Fig_tsigma} and \ref{Fig_thres}, where results with and
without a density dependent shift of the $\omega$ pole mass for the
targets \nuc{12}{C} and \nuc{208}{Pb} are shown. The real part of
the $\omega$ in-medium self energy we parametrize by the attractive
mass shift as given in Eq.~(\ref{mass}) with the canonical strength
parameter $\alpha=0.16$ \cite{Hatsuda:1991ez,Klingl:1997kf}, that is
also consistent with the measurement of the in-medium $\omega$ mass
reported in \cite{Trnka:2005ey}. At low photon
energy the cross section including the dropping $\omega$ mass
shows a pronounced excess over the
standard calculation due to the lowering of the $\omega N$ threshold
in the medium. However, the transparency ratio shows an enhancement
towards the $\omega N$ threshold\footnote{Also the diminishing
recoil due to coherent $\omega$ production would lead to a shift of
the threshold with the mass number $A$. This shift, however, does
not show up in the quasielastic events considered here.} already
without any real $\omega$ potential due to the in-medium broadening
of the $\omega$ spectral function and the energy smearing caused by
Fermi motion. The threshold enhancement is magnified when the
attractive mass shift according to Eq.~(\ref{mass}) is turned on.
The production of $\omega$ mesons at energies below the free
production threshold is dominated by the low-energy tail of the
$\omega$ in-medium spectral function. Such $\omega$ mesons far
off-shell the free $\omega$ mass are difficult to identify
experimentally, but they do contribute to the total $\omega$ yield
and thus influence the transparency ratio. Therefore, we again show
results with the condition $0.75~\mathrm{GeV}\le
M\le0.81~\mathrm{GeV}$. Imposing this restriction on the
$\pi^0\gamma$ mass, the component of low-mass $\omega$ mesons in the
final particle yield is discarded. That the dotted curve in Fig.
\ref{Fig_thres} shows a strong rise at small energies is due to the
fact that here the mass-cut influences the elementary cross section
even more than the nuclear one. Cutting out the low mass tails of
the spectral function of the free $\omega$ leads to the drastic
rise, which is thus not an in-medium effect.

However strong the threshold enhancement shows up, any extraction of
the real part of the in-medium self energy has to relie on the solid
line in Fig. \ref{Fig_thres} as a baseline. This, however, is
extremely sensitive to the theoretical assumptions for the $\omega$
photoproduction cross section. Thus, a complete understanding of the
elementary production process is required. Theoretical efforts
towards such a complete description of $\omega$ photoproduction from
elementary targets are, however, underway \cite{Shklyar:2004ba}. At
present, a unique mapping of the observed effects to particular
medium modifications turns out to be difficult as the relevant
contributions to the $\omega$ photoproduction process have not yet
been resolved.

\subsection{Medium modifications of the $\omega$ decay width}

As for the $\omega$ also the spectral function of the $\rho$
meson is expected to change in the nuclear medium
\cite{Klingl:1997kf,Peters:1997,Post:2000qi,Post:2003hu}. Such a
modification of the in-medium $\rho$ spectrum has also an impact on
the $\omega$ self energy as the most important decay channel of the
$\omega$ in vacuum is $\omega\rightarrow3\pi$ that is dominated by
the Gell-Mann-Sharp-Wagner (GSW) process, a process where the
$\omega$ first converts into an intermediate $\rho\pi$ state
followed by the decay of the virtual $\rho$ into two pions. The
$\omega\rightarrow 3\pi$ decay width then is given by
\begin{eqnarray}\label{omedec}
\Gamma_{\omega\rightarrow3\pi}(s,\rho)=\frac{3g^2}{4\pi m_{\pi}^2}
\int\limits_{4m_{\pi}^2}^{(\sqrt{s}-m_{\pi}^2)^2}dm_{\rho}^2~
q^3\mathcal{A}_{\rho}(m_{\rho},\rho)
\frac{\Gamma(\rho\rightarrow\pi\pi)}{\Gamma_{\mathrm{tot}}(m_{\rho},\rho)},
\end{eqnarray}
where $\mathcal{A}_{\rho}$ is the (in-medium) spectral function of
the $\rho$ meson and $\Gamma_{\mathrm{tot}}$ is its total
(in-medium) width. The $\omega-\rho\pi$ coupling
constant $g$ is determined by the postulate that at the on-shell
point the experimental width of 7.5 MeV \cite{Eidelman:2004wy} is
obtained. The phase-space factor $q$ is given by the center-of-mass
momentum of the virtual $\rho$ meson and pion:
\begin{eqnarray}
q\equiv p_f(s,m_{\rho},m_{\pi})=
\sqrt{\frac{1}{4s}\left((s-m_{\rho}^2-m_{\pi}^2)^2-4m_{\rho}^2m_{\pi}^2\right)},
\end{eqnarray}
where $s$ is the mass of decaying $\omega$ meson squared and
$m_{\rho}$ is the mass of the virtual $\rho$ meson. For the vacuum
spectral density of the $\rho$ meson we take its coupling to a $2\pi$
state into account. The $\rho$ decay width to this channel
is given by \cite{Klingl:1996by}
\begin{eqnarray}
\Gamma(\rho\rightarrow\pi\pi)=\frac{f_{\rho}^2}{48\pi}
m_{\rho}\left[1-4\frac{m_{\pi}^2}{m_{\rho}^2}\right]^{\frac{3}{2}},
\end{eqnarray}
where again the coupling is obtained from the experimental $\rho$
width of 149.2 MeV. Due to the dependence of the phase-space factor
$q^3$ on $m_\rho$ only the low-energy tail of the $\rho$ spectral
function contributes to the integral in Eq.~(\ref{omedec}).
Therefore the $\omega\rightarrow3\pi$ width increases dramatically
for higher $\omega$ masses as more and more of the $\rho$ strength
is picked up by the integral.
\begin{figure}
\begin{center}
\includegraphics[scale=1.0]{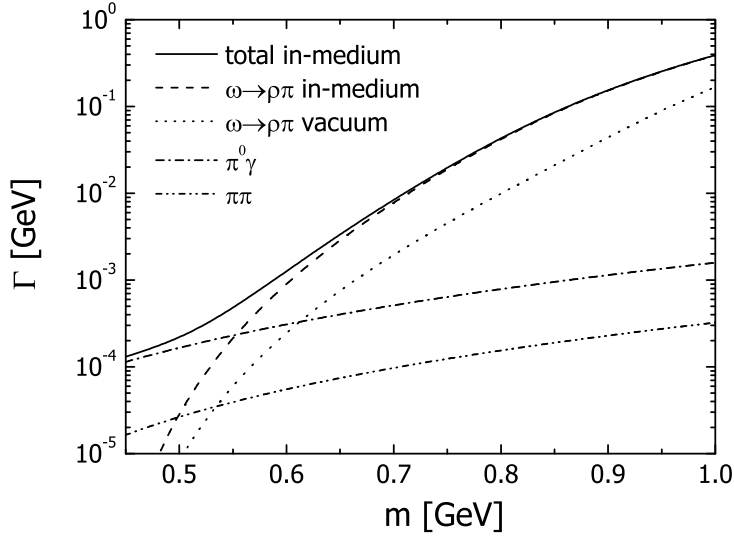}
\caption{Partial and total decay widths of the $\omega$ meson
in vacuum and at normal nuclear matter density. The $\omega$ 
collisional width is not included. At the on-shell
point the $\rho\pi$ decay width amounts to 7.5 MeV in vacuum, whereas at
$\rho_0$ it increases to 31.8 MeV.} \label{Fig_width}
\end{center}
\end{figure}

Going to the nuclear medium, the $\rho$ spectral distribution
is broadened due to elastic and inelastic $\rho N$ collisions
that lead to a shorter lifetime of the interacting $\rho$
state. We include this broadening by adding the phenomenological
collisional width of roughly 100 MeV at normal nuclear
matter density \cite{Effenberger:1999ay} to the total
$\rho$ width in the medium:
\begin{eqnarray}
\Gamma_{\mathrm{tot}}(m_{\rho},\rho)=\Gamma(\rho\rightarrow\pi\pi)
+0.1~\mathrm{GeV}\cdot\frac{\rho}{\rho_0}.
\end{eqnarray}
On top of that we also consider a dropping $\rho$ mass analogous to
the dropping $\omega$ mass in matter given by Eq.~(\ref{mass}). We
note that not only a shift of the $\rho$ pole mass but also a very
general reshuffling of spectral strength to the low energy part of
the $\rho$ spectral function would result in a similar effect on the
$\omega$ decay width. Such a shift of strength could be
caused by the excitation of subthreshold nucleon resonances as
for instance obtained in the sophisticated approach of
Ref.~\cite{Post:2003hu}. Our result at normal nuclear matter density 
excluding the $\omega$ collisional width is
shown in Fig.~\ref{Fig_width}. Due to the fact that more of the
spectral strength of the $\rho$ meson lies inside the bounds of the
integration in Eq.~(\ref{omedec}), the $\rho\pi$ width of the
$\omega$ increases from 7.5 MeV in vacuum to 31.8 MeV at normal
nuclear matter density.

\begin{figure}
\begin{center}
\includegraphics[scale=1.]{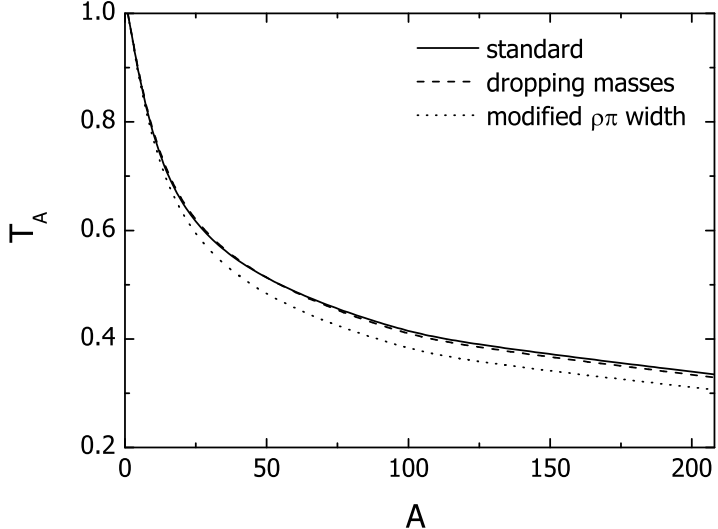}
\caption{Nuclear transparency ratio at $E_\gamma = 1.5$ GeV. The
dashed and dotted lines include dropping vector meson masses. For
the dotted line in addition the modification of the
$\omega\rightarrow\rho\pi$ decay width has also been taken into
account.} \label{Fig_ratio2}
\end{center}
\end{figure}

In Fig.~\ref{Fig_ratio2} we show the nuclear transparency with and
without including the modified $\omega\rightarrow\rho\pi$ decay 
width on top of the $\omega$
collisional width. The dropping $\rho$ and $\omega$ masses alone
have no effect on the transparency ratio as the photon energy is
well above threshold where the change in the phase space factors
becomes small. Including the modified $\rho\pi$ decay width, more
$\omega$ mesons decay to that channel due to the opening of the
$\rho\pi$ phase space at non-zero nuclear density. Hence more
$\omega$ mesons are taken out of the total flux inside the nucleus.
This leads to a further but small reduction of the transparency ratio.

\subsection{Sensitivity to the $\omega N$ absorption cross section}

\begin{figure}
\begin{center}
\includegraphics[scale=1.]{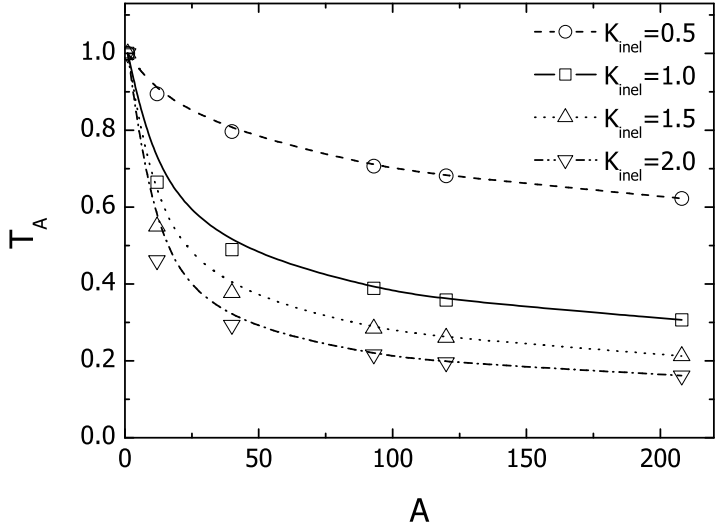}
\caption{Nuclear transparency ratio obtained from BUU transport
simulations as function of the target mass number. Calculations
have been done for different inelastic $K$-factors. The photon energy is
$E_{\gamma}=1.5$ GeV. No acceptance corrections have been
applied.}\label{Fig_ratio3}
\end{center}
\end{figure}

In view of the uncertainties connected with the
$\omega N$ cross section we show in Fig.~\ref{Fig_ratio3} our
results for the transparency ratio using different assumptions for
the $\omega$ collisional width. To this end we multiply the
collisional part of the $\omega$ self energy with a constant
normalization $K$-factor. According to the low density theorem
(\ref{gcoll}), a modification of the collisional width goes along
with an analogous change of the total $\omega N$ cross section, i.~e.
\begin{eqnarray}
\tilde\Gamma_{\mathrm{coll}}=K_{\mathrm{inel}}\cdot\Gamma_{\mathrm{coll}}
\leftrightarrow\tilde\sigma_{VN}=K_{\mathrm{inel}}\cdot\sigma_{VN}~.
\end{eqnarray}
In the present calculations, we put this modification entirely
into the absorptive part of the $\omega$ self energy, i.~e.
\begin{eqnarray}
\tilde\sigma_{VN}^{\mathrm{inel}} &=& K_{\mathrm{inel}}\cdot\sigma_{VN}^{\mathrm{tot}}-
\sigma_{VN}^{\mathrm{inel}}\\
\tilde\sigma_{VN}^{\mathrm{el}} &=& \sigma_{VN}^{\mathrm{el}}~,
\end{eqnarray}
where $\sigma_{\mathrm{tot}}$ is the sum of the elastic and
inelastic channels. After all, the transparency ratio without any
restrictions on angles and momentum of the produced particles will
be sensitive primarily to the absorptive part of the $\omega N$
interaction as quasi elastic scattering processes do not lead to a
loss of flux.

Besides the results obtained with the canonical value for the
$\omega$ in medium width ($K_{\mathrm{inel}}=1.0$, $\Gamma_{\mathrm{coll}}=37$
MeV), curves with $K_{\mathrm{inel}}=0.5$, $K_{\mathrm{inel}}=1.5$
 and $K_{\mathrm{inel}}=2.0$ are shown. We
also include the dropping $\rho$ and $\omega$ masses as well as
the modified $\rho\pi$ decay width as discussed previously. The
results from our transport calculations show an obvious lowering
of the nuclear transparency ratio as the $\omega$ width goes up
and, hence, the mean free path of the $\omega$ shrinks to smaller
values. The transparency ratio decreases in a non-linear way with
the increasing $\omega$ width.
Whereas the absolute size of the transparency
ratio yields important information about the $\omega$ collisional
width, its $A$-($N$-, $Z$-) scaling in principle is also sensitive
to the isospin dependence of the production and absorption cross
sections.

\subsection{Elastic scattering}

Finally, we explore the influence of different assumptions
for the $\omega N$ elastic scattering cross section on the 
nuclear transparency. Again we use a constant normalization
factor $K_{\mathrm{el}}$ that we now multiply to the elastic 
scattering cross section
\begin{eqnarray}
\tilde\sigma_{VN}^{\mathrm{el}}
&=&K_{\mathrm{el}}\cdot\sigma_{VN}^{\mathrm{el}}\\
\tilde\sigma_{VN}^{\mathrm{tot}}
&=&\tilde\sigma_{VN}^{\mathrm{el}}
+\sigma_{VN}^{\mathrm{inel}}.
\end{eqnarray}
As said earlier, elastic scattering processes do not lead to a
loss of flux and, thus, do not directly influence the total
nuclear cross section and the transparency. On the other hand,
such scattering processes change the $\omega$ momentum distribution
and, therefore, help to keep the $\omega$ mesons inside the medium
for a longer time. The stopping of the $\omega$ mesons in nuclear 
matter due to elastic $\omega N$ scattering is particularly large 
as the mass of the $\omega$ is comparable to the nucleon mass, what 
leads to a relatively high energy loss of the $\omega$ in these
collisions.

\begin{figure}
\begin{center}
\includegraphics[scale=1.]{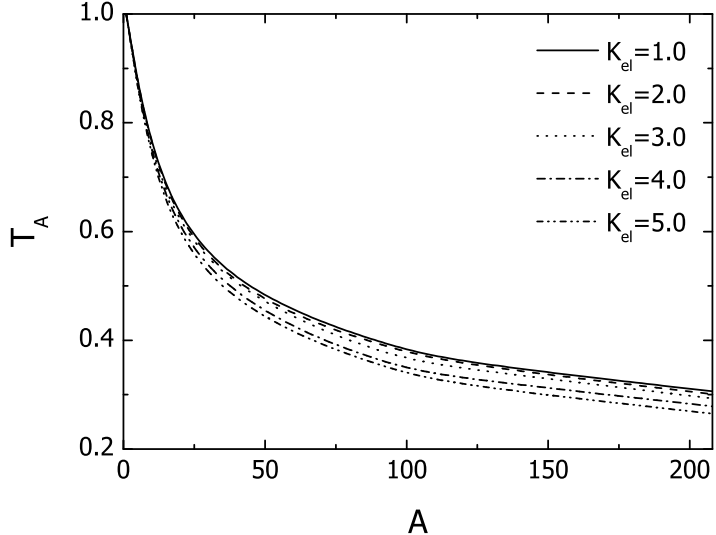}
\caption{Nuclear transparency ratio at $E_{\gamma}=1.5$ GeV. 
Calculations for different elastic $K$-factors from
$K_{\mathrm{el}}=1.0$ to $K_{\mathrm{el}}=5.0$ are shown.}
\label{Fig_ratio4}
\end{center}
\end{figure}

\begin{figure}
\begin{center}
\includegraphics[scale=1.]{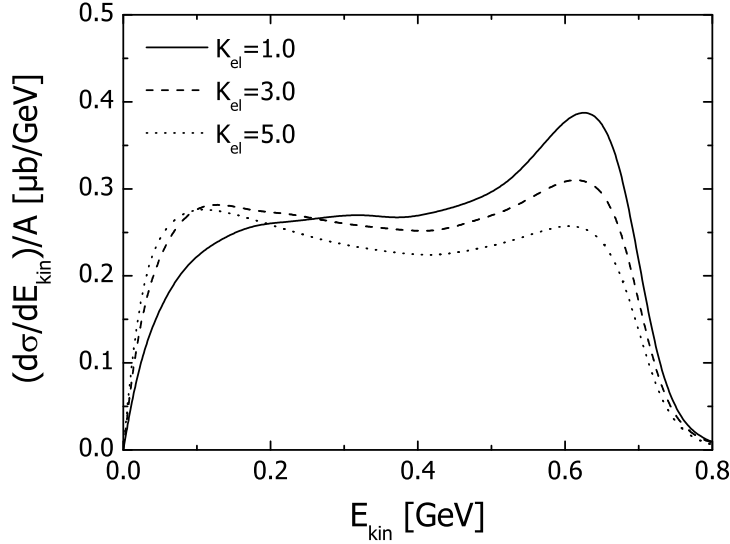}
\caption{Kinetic energy spectra of $\pi^0\gamma$ pairs from \nuc{208}{Pb}
at $E_{\gamma}=1.5$ GeV.}
\label{Fig_kinetic}
\end{center}
\end{figure}

The transparency ratio with normalization factors $K_{\mathrm{el}}=1,2,3,4$ 
and $5$ is shown in Fig.~\ref{Fig_ratio4}. The impact of the different 
elastic $K$-factors indeed is small. In Fig.~\ref{Fig_kinetic} we also show
the kinetic energy spectra resulting from these calculations, here for
the case of the \nuc{208}{Pb} target. The peak at high kinetic energies
is due to the forward peaked angular distribution for the $N(\gamma,\omega)N$
process. Turning on elastic $\omega N$ scattering, $\omega$ mesons from
the high energy part of the spectrum are shuffled to the low energy tail.
On one hand, these $\omega$ mesons stay in the medium for a longer time
where they have the chance to get absorbed in inelastic $\omega N$ collisions.
On the other hand, the $\omega N$ absorption cross sections are particularly
large at low energies due to the open phase space at the $\omega N$ threshold
for processes like $\omega N\rightarrow\pi N$, $\omega N\rightarrow\pi\pi N$ etc.
Hence, we observe a slight reduction of the nuclear transparency with the
increasing elastic $\omega N$ cross section. Note, that the effect on the 
transparency ratio becomes significant only for $K$-factors $K_{\mathrm{el}}=4,5$
what means a rather drastic change of the elastic scattering cross section as
compared to our standard estimate Eq.~(\ref{sel}).

\subsection{Final remarks}

At first glance it appears to be quite astonishing that the enhanced
collisional width has a much larger impact on the transparency than
a modification of the $\omega\rightarrow\rho\pi$ decay channel. The
reason for this effect lies in the momentum dependence of both
contributions to the in-medium self energy. In our model the
$\omega$ collisional width is -- except for the threshold region
where the nucleon momentum distribution plays an important role --
proportional to the $\omega$ momentum with respect to the nuclear
medium. Thus the collisional width of roughly 37 MeV at normal
nuclear matter density for $\omega$ mesons at rest blows up to
almost 150 MeV for a typical laboratory momentum of 1 GeV. In contrast, the
modification of the $\rho\pi$ decay width generated by the purely
density-dependent $\rho$ meson potential is independent of the
$\omega$ momentum. In the considered photon energy regime, the
$\omega$ momentum distribution is peaked around $p=1$ GeV. Hence, a
doubling of the inelastic collisional self energy 
($K_{\mathrm{inel}}=1\rightarrow K_{\mathrm{inel}}=2$) has
an effect on the $\omega$ attenuation that is roughly one order of
magnitude larger than the impact of the decreasing $\rho$ meson
mass.

In fact, the nuclear transparency is sensitive not only to the
$\omega N$ interaction in nuclei but also to medium modifications of
the production processes and decay widths of the $\omega$ meson. In
particular, also the $\omega$ production rates from neutrons have to
be known in order to obtain precise information on the $\omega$
nucleus potential. In view of that, it might be desirable to
normalize the nuclear transparency ratio not to the proton cross
section but to the $\omega$ yield obtained from Deuterium or Carbon
targets. Further uncertainties concerning the determination of the
real as well as the imaginary part of the $\omega$ nucleus potential
arise from the possibility that also the $\pi^0\gamma$ decay width
as well as other decay channels might become modified in a
surrounding with non-zero nuclear density. In particular the change
of the $\rho$ spectral function in nuclei and its back coupling on
the $\omega$ in-medium self energy are still open issues.
Experimentally, it is essential to understand the background so that
the actual production cross section (that is not necessarily
required for a verification of a shape change of the $\omega$
in-medium spectral function) and, correspondingly, the transparency
are quantitatively reliable.


\section{Glauber approximation}\label{glauber}

In this section we will finally discuss $\omega$ photoproduction
in the semi-analytic Glauber picture as
a much simpler means to extract the inelastic $\omega N$
cross section from the total photoproduction cross section. After
our detailed study of various nuclear effects within the coupled
channel transport approach, we have found that the nuclear
transparency ratio at photon energies well above threshold is first
of all sensitive to the $\omega N$ absorption cross section whereas
other medium effects such as dropping vector meson masses and the
modified decay width give rise to only small corrections. We note,
however, that this statement holds only as long as total inclusive
observables without any cuts on angle and momentum of the observed
vector mesons are considered. A limited experimental acceptance
possibly introduces dependences on elastic scattering processes and
details of the dynamics that cannot be included in the semi-analytic
Glauber framework.

\subsection{Analytic expressions}

In the Glauber-eikonal approximation, neglecting Fermi motion,
Pauli blocking, coupled-channel effects, nuclear shadowing and 
quasi elastic scattering processes, the incoherent single meson 
photoproduction cross section has the following form 
\cite{Muehlich:2005kf}:
\begin{eqnarray}\label{sigma}
\sigma_{\gamma A} &=& \int d^3r\rho({\bf r})\sigma_{\gamma N}
\exp\left[-\sigma^{\rm
inel}_{VN}\int\limits_{z}^{\infty}dz'\rho({\bf b},z')\right] ~.
\end{eqnarray}
where $\rho({\bf r})$ is the nuclear density distribution, $\sigma_{\gamma N}$
the total vector meson photoproduction cross section on a single nucleon and 
$\sigma_{VN}^{\mathrm{inel}}$ ($\equiv\sigma_{VN}$ in the following) 
the vector meson nucleon absorption cross section.
In Eq.~(\ref{sigma}) we have neglected any $\pi$ final state interactions which,
anyway, give only marginal corrections to the nuclear cross section as most 
of the $\omega$ mesons decay outside the nucleus.

In order to carry out the integrals explicitly we
approximate the nuclear density distribution by a sphere that is
filled up homogeneously with $A$ nucleons
\begin{eqnarray}
\rho({\bf r})=\rho_0\Theta(|{\bf r}|-R)=\frac{3A}{4\pi R^3}\Theta(|{\bf r}|-R),
\end{eqnarray}
where $R$ is the nuclear radius that we parametrize according to
$R=r_0\cdot A^{1/3}$. For the radius parameter $r_0$ we use the
numerical value $r_0=1.143~\mathrm{fm}$ in order to be consistent
with $\rho_0=0.16~\mathrm{fm}^{-3}$. Going to cylindrical
coordinates and assuming that the elementary cross section does not
depend on the density, we are then able to rewrite Eq.~(\ref{sigma})
in the following way:
\begin{eqnarray}\label{glaub1}
\sigma_{\gamma A}=\int d^2b\, dz\,
\frac{\sigma_{\gamma N}}{\sigma_{VN}}\frac{\partial}{\partial z}
\exp\left[-\sigma_{VN}\int\limits_z^{\infty}dz'\rho({\bf
b},z')\right]
\end{eqnarray}
where now $\bf b$ is the 2-dimensional coordinate in the plane
perpendicular to the incoming photon direction. At this stage we
introduce one further abbreviation, namely
\begin{eqnarray}
\lambda_0=\frac{1}{\sigma_{VN}\rho_0}
\end{eqnarray}
which is the mean free path of the vector meson at normal nuclear
matter density with respect to inelastic vector meson nucleon
collisions. Evaluating Eq.~(\ref{glaub1}) further we obtain
\begin{eqnarray}
\sigma_{\gamma A} &=& 2\pi\frac{\sigma_{\gamma N}}{\sigma_{VN}}
\int\limits_0^R b\,db\,\left\{1-\exp\left[-\frac{1}{\lambda_0}
\int\limits_{-\infty}^{+\infty}dz'\Theta(\sqrt{b^2+z'^2}-R)\right]\right\}\\
&=& \pi R^2\frac{\sigma_{\gamma N}}{\sigma_{VN}}\left\{1-\frac{2}{R^2}
\int\limits_0^R bdb\exp\left[-2\frac{\sqrt{R^2-b^2}}{\lambda_0}\right]\right\}.
\end{eqnarray}
Carrying out the remaining integral, we finally obtain the following
expression for the nuclear transparency ratio:
\begin{eqnarray}\label{glaub2}
T_A &=& \frac{\pi R^2}{A\sigma_{VN}}\nonumber\\
&&\times\left\{1+\left(\frac{\lambda_0}{R}\right)
\exp\left[-2\frac{R}{\lambda_0}\right]+
\frac{1}{2}\left(\frac{\lambda_0}{R}\right)^2
\left(\exp\left[-2\frac{R}{\lambda_0}\right]-1\right)\right\}.
\end{eqnarray}
In the limit $\sigma_{VN} \to \infty$, i.~e. $\lambda_0/R \to 0$ this
just reduces to
\begin{equation}\label{Gllow}
T_A \longrightarrow \frac{\pi R^2}{A \sigma_{VN}}~.
\end{equation}

\begin{figure}
\begin{center}
\includegraphics[scale=1.]{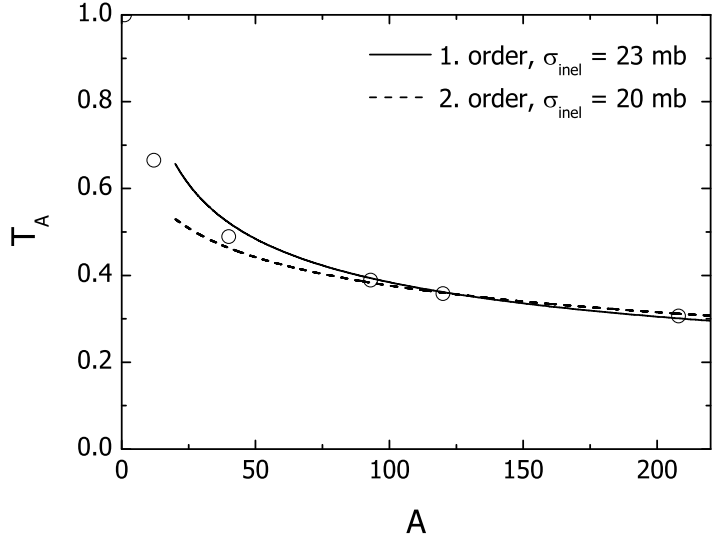}
\caption{Nuclear transparency ratio according to Eq.~(\ref{glaub2})
fitted to BUU calculations (open circles) from Fig.~\ref{Fig_ratio3} with
$K_{\mathrm{inel}}=K_{\mathrm{el}}=1$. For details see text.}
\label{Fig_fit}
\end{center}
\end{figure}

\subsection{Fit results}

We now fit these expressions, using the inelastic vector meson
nucleon cross section $\sigma_{VN}$ as an open parameter, to our BUU
calculations shown in Fig.~\ref{Fig_ratio3} with $K_{\mathrm{inel}}=K_{\mathrm {el}}=1$.
To this end we include only nuclei starting from \nuc{40}{Ca} since our starting
point, i.~e. the Glauber expression for the nuclear photoproduction
cross section Eq.~(\ref{glaub1}), is valid only for large nuclei
with a relative error of the order of $A^{-1}$
\cite{Glauber:1959,Glauber:1970}. The results of this procedure are
shown in Fig.~\ref{Fig_fit}. Adopting the lowest order expression
(\ref{Gllow}) we find an inelastic $\omega N$ cross section of 23
mb, whereas including the correction factor in the curled brackets
in Eq.~(\ref{glaub2}) an inelastic cross section of 20 mb is found.
Hence, the correction terms are only of minor importance. This is
particularly true for large nuclei as the additional terms come with
powers of $(\lambda_0/R)$ which is a small parameter for heavy
nuclei and large absorption cross sections. 

As said earlier, the
momentum distribution of the $\omega$ mesons in the data sample is
peaked around $p_V=1$~GeV. Hence, the inelastic $\omega N$ cross
section in our transport calculations with $K=1$ amounts to roughly
24 mb, see Eq.~(\ref{sinel}). This is in astonishing agreement with
the much simpler Glauber result. The somewhat smaller values
extracted by means of the analytic formula (\ref{glaub2}) can be
attributed to the disregard of any target surface region where the mean
free path of the vector mesons again becomes large, giving rise to a
diminishing absorption probability for any produced particle on its
way out of the nucleus. We conclude that the Glauber model provides
a quite reliable tool to extract the $\omega N$ inelastic cross
section. Actual experiments, however, often have geometrical and
kinematical acceptance limitations which then require the full
transport calculation as we have shown for the case of $\phi$ 
photoproduction in \cite{Muehlich:2005kf}.

\section{Summary}\label{Summary}

In summary, we have shown that a measurement of the nuclear
transparency ratio can indeed yield important information on the
real and imaginary part of the $\omega$ nucleus potential. To this
end we have included all nuclear effects in a coupled-channel
transport calculation that describes the incoherent contribution to
photon-nucleus reactions. The imaginary part of the $\omega$
in-medium self energy can be extracted from the $A$-dependence of
the nuclear transparency ratio by fitting the total inelastic
$\omega N$ cross section to experimental data. For such a
measurement we consider a photon beam energy of around 1.5 GeV as
optimal. Studying total inclusive observables only, the inelastic
vector meson nucleon cross section can also be extracted from the
naive Glauber multiple scattering result with a relative error on
the $10-20\%$ level.

Also the real part of the $\omega$ in-medium self energy can at
least in principle be studied from an examination of the energy
dependence of the transparency ratio in the threshold region.
Uncertainties arise from the fact that not only the intrinsic
properties of the $\omega$ meson but also the $\omega$ production
cross section or its decay width can experience modifications from
the interaction of the involved particles with the surrounding
nuclear matter. Any statement about the real part of the $\omega$
nucleus potential has to relie on assumptions for the elementary
$\omega$ production cross section. It is also important to realize
that the attenuation measurement as considered in the analysis at
hand is sensitive not to the $\omega$ self energy at rest but to the
$\omega$ properties at rather high momenta, i.~e. the momentum range
around 1 GeV.

The authors gratefully acknowledge stimulating discussions with
D.~Trnka and V.~Metag on the subject.



\begin{thebibliography}{ccc}

\bibitem{Agakichiev:2005ai}
  G.~Agakichiev {\it et al.}  [CERES Collaboration],
  Eur.\ Phys.\ J.\ C {\bf 41} (2005) 475.

\bibitem{Usai:2005zh}
  G.~Usai {\it et al.}  [NA60 Collaboration],
  Eur.\ Phys.\ J.\ C {\bf 43} (2005) 415.

\bibitem{Adams:2003cc}
  J.~Adams {\it et al.}  [STAR Collaboration],
  Phys.\ Rev.\ Lett.\  {\bf 92} (2004) 092301.

\bibitem{Trnka:2005ey}
  D.~Trnka {\it et al.}  [CBELSA/TAPS Collaboration],
  Phys.\ Rev.\ Lett.\  {\bf 94} (2005) 192303 .

\bibitem{Ozawa:2000iw}
  K.~Ozawa {\it et al.}  [E325 Collaboration],
  Phys.\ Rev.\ Lett.\  {\bf 86} (2001) 5019.

\bibitem{Brown:1991kk}
  G.~E.~Brown and M.~Rho,
  Phys.\ Rev.\ Lett.\  {\bf 66} (1991) 2720.
  
\bibitem{Hatsuda:1991ez}
  T.~Hatsuda and S.~H.~Lee,
  Phys.\ Rev.\ C {\bf 46} (1992) 34.

\bibitem{Leupold}
  S.~Leupold, W.~Peters and U.~Mosel,
  Nucl.\ Phys.\ A {\bf 628} (1998) 311.
  
\bibitem{Bernard:1988db}
  V.~Bernard and U.~G.~Meissner,
  Nucl.\ Phys.\ A {\bf 489} (1988) 647.

\bibitem{Klingl:1997kf}
  F.~Klingl, N.~Kaiser and W.~Weise,
  Nucl.\ Phys.\ A {\bf 624} (1997) 527.

\bibitem{Klingl:1998zj}
  F.~Klingl, T.~Waas and W.~Weise,
  Nucl.\ Phys.\ A {\bf 650} (1999) 299.

\bibitem{Post:2000rf}
  M.~Post and U.~Mosel,
  Nucl.\ Phys.\ A {\bf 688} (2001) 808.

\bibitem{Lutz:2001mi}
  M.~F.~M.~Lutz, G.~Wolf and B.~Friman,
  Nucl.\ Phys.\ A {\bf 706} (2002) 431.

\bibitem{Klingl:1996by}
  F.~Klingl, N.~Kaiser and W.~Weise,
  Z.\ Phys.\ A {\bf 356} (1996) 193.

\bibitem{Muhlich:2003tj}
  P.~Muhlich, T.~Falter and U.~Mosel,
  Eur.\ Phys.\ J.\ A {\bf 20} (2004) 499.

\bibitem{Djalali:2004}
  C.~Djalali {\it et al.} [CLAS collaboration],
  private communication (2004).

\bibitem{Alvensleben:1970uw}
  H.~Alvensleben {\it et al.},
  Phys.\ Rev.\ Lett.\  {\bf 24} (1970) 786.

\bibitem{Cabrera:2003wb}
  D.~Cabrera, L.~Roca, E.~Oset, H.~Toki and M.~J.~Vicente Vacas,
  Nucl.\ Phys.\ A {\bf 733} (2004) 130.

\bibitem{Muehlich:2005kf}
  P.~Muehlich and U.~Mosel, Nucl.\ Phys.\ A {\bf 765} (2005) 188.

\bibitem{Ishikawa:2005aw}
  T.~Ishikawa {\it et al.},
  Phys.\ Lett.\ B {\bf 608} (2005) 215.

\bibitem{Effenberger:1999ay}
  M.~Effenberger, E.~L.~Bratkovskaya and U.~Mosel,
  Phys.\ Rev.\ C {\bf 60} (1999) 044614.

\bibitem{Muhlich:2002tu}
  P.~Muhlich, T.~Falter, C.~Greiner, J.~Lehr, M.~Post and U.~Mosel,
  Phys.\ Rev.\ C {\bf 67} (2003) 024605.

\bibitem{Andersson:1992iq}
  B.~Andersson, G.~Gustafson and H.~Pi,
  Z.\ Phys.\ C {\bf 57} (1993) 485.

\bibitem{Teis:1996kx}
  S.~Teis, W.~Cassing, M.~Effenberger, A.~Hombach, U.~Mosel and G.~Wolf,
  Z.\ Phys.\ A {\bf 356} (1997) 421.

\bibitem{Lehr:1999zr}
  J.~Lehr, M.~Effenberger and U.~Mosel,
  Nucl.\ Phys.\ A {\bf 671} (2000) 503.

\bibitem{Falter:2002jc}
  T.~Falter and U.~Mosel,
  Phys.\ Rev.\ C {\bf 66} (2002) 024608.

\bibitem{Penner:2002md}
  G.~Penner and U.~Mosel,
  Phys.\ Rev.\ C {\bf 66} (2002) 055212.

\bibitem{Barth:2003}
  J.~Barth {\it et al.} [SAPHIR collaboration],
  Eur.\ Phys.\ J.\ A {\bf 18} (2003) 117.

\bibitem{Lykasov:1998ma}
  G.~I.~Lykasov, W.~Cassing, A.~Sibirtsev and M.~V.~Rzyanin,
  Eur.\ Phys.\ J.\ A {\bf 6} (1999) 71.

\bibitem{Shklyar:2004ba}
  V.~Shklyar, H.~Lenske, U.~Mosel and G.~Penner,
  Phys.\ Rev.\ C {\bf 71} (2005) 055206
  [Erratum-ibid.\ C {\bf 72} (2005) 019903].

\bibitem{Messchendorp:2001pa}
  J.~G.~Messchendorp, A.~Sibirtsev, W.~Cassing, V.~Metag and S.~Schadmand,
  Eur.\ Phys.\ J.\ A {\bf 11} (2001) 95.
  
\bibitem{Peters:1997}
  W.~Peters, M.~Post, H.~Lenske, S.~Leupold and U.~Mosel,
  Nucl.\ Phys.\ A {\bf 632} (1998) 109.

\bibitem{Post:2000qi}
  M.~Post, S.~Leupold and U.~Mosel,
  Nucl.\ Phys.\ A {\bf 689} (2001) 753.

\bibitem{Post:2003hu}
  M.~Post, S.~Leupold and U.~Mosel,
  Nucl.\ Phys.\ A {\bf 741} (2004) 81.

\bibitem{Eidelman:2004wy}
  S.~Eidelman {\it et al.}  [Particle Data Group],
  Phys.\ Lett.\ B {\bf 592} (2004) 1.

\bibitem{Glauber:1959}
  R.~J.~Glauber
  {\it "High energy collision theory"},
  in {\it "Lectures in Theoretical Physics Vol.~1"},
  Wiley Intersience, New York (1959) 315.

\bibitem{Glauber:1970}
  R.~J.~Glauber
  {\it "Theory of high energy hadron-nucleus collisions"},
  in {\it "High energy physics and nuclear structure"},
  Plenum, New York (1970) 207.

\end{thebibliography}
\end{document}